# The CMS Magnetic Field Map Performance

V. I. Klyukhin, N. Amapane, V. Andreev, A. Ball, B. Curé, A. Hervé, A. Gaddi, H. Gerwig, V. Karimaki,
R. Loveless, M. Mulders, S. Popescu, L. I. Sarycheva, and T. Virdee

*Abstract*—The Compact Muon Solenoid (CMS) is a general-purpose detector designed to run at the highest luminosity at the CERN Large Hadron Collider (LHC). Its distinctive features include a 4 T superconducting solenoid with 6 m diameter by 12.5 m long free bore, enclosed inside a 10000-ton return yoke made of construction steel. Accurate characterization of the magnetic field everywhere in the CMS detector is required. During two major tests of the CMS magnet the magnetic flux density was measured inside the coil in a cylinder of 3.448 m diameter and 7 m length with a specially designed field-mapping pneumatic machine as well as in 140 discrete regions of the CMS yoke with NMR probes, 3-D Hall sensors and flux-loops. A TOSCA 3-D model of the CMS magnet has been developed to describe the magnetic field everywhere outside the tracking volume measured with the field-mapping machine. A volume based representation of the magnetic field is used to provide the CMS simulation and reconstruction software with the magnetic field values. The value of the field at a given point of a volume is obtained by interpolation from a regular grid of values resulting from a TOSCA calculation or, when available, from a parameterization. The results of the measurements and calculations are presented, compared and discussed.

*Index Terms*—Flux-loops, hall probes, magnetic field measurements, NMR probes, superconducting solenoid.

## I. INTRODUCTION

THE magnetic system of the Compact Muon Solenoid (CMS) general-purpose detector at the CERN Large Hadron Collider (LHC) consists of a 4 T superconducting coil with 6 m diameter by 12.5 m long free bore and a 10000-ton yoke [1], [2]. The yoke is assembled with the construction steel plates up to 630 mm thick which return the flux of the superconducting solenoid, serve as the absorber plates of the muon detection system, and provide additional bending power for a measurement of the muon momentum independent of the inner tracking system [3]. The yoke includes five dodecagonal three-layered barrel wheels, four end-cap disks at each end, and the ferromagnetic parts of forward hadron calorimeter and shield of the LHC magnets.

The continuous direct measurements of the magnetic flux density $B$ have been performed in a cylinder of 1.724 m radius and 7 m long inside the CMS barrel hadron calorimeter where the electromagnetic calorimeter and inner tracker are located. The measurements of all three components of $B$ are done with a pneumo-mechanical fieldmapper designed and produced at Fermilab [4]. The fieldmapper used 10 3-D B-sensors (Hall probes) developed at NIKHEF and calibrated at CERN to precision $5 \cdot 10^{-4}$ at 4.5 T field [5]. This device scanned the cylinder volume with an increment of 0.05 m along the CMS coil axis and measured $B$ rotating two arms with B-sensors by increments of $7.5°$ in azimuth angle in 141 vertical planes orthogonal to the coil axis. These precise measurements are accomplished at five different values of the magnetic flux density in the center of the superconducting solenoid, $B_0$: 2.02, 3.02, 3.52, 3.81, and 4.01 T. At $B_0$ of 4.01 T the magnetic flux density is measured with two NMR probes moved with the fieldmapper along the coil axis and along the maximum radius of 1.724 m in horizontal plane.

Four NMR probes monitor $B$ inside the superconducting solenoid at radii of 2.91 m near the coil transverse middle plane. Two NMR probes measure the magnetic flux density near the coil axis at the axial extremes of the inner tracker.

The continuous direct measurements of $B$ outside the CMS coil are extremely difficult to perform. To measure the magnetic flux density in the discrete points of air gaps between the steel yoke elements, 91 immovable and 7 movable B-sensors calibrated at CERN to precision better than $10^{-3}$ at 1.4 T field have been used.

To estimate the magnetic flux density in 22 cross sections of the yoke, flux-loops of $315 \div 495$ turns have been wound around selected CMS yoke plates. The measurements of $B$ in steel [6] are performed using the magnetic flux changes induced by the "fast" (190 s time-constant) discharges of the CMS coil made possible by the protection system, which is provided to protect the magnet in the event of major faults [7], [8]. An integration technique [9], [10] was developed to reconstruct the average initial magnetic flux density in steel blocks at the full magnet excitations, and the contribution of the eddy currents was calculated with ELECTRA [11] and estimated on the level of a few per cent [12].

The description of the magnetic flux density distribution in full detector volume is done with a 3-D model of the CMS magnetic system [13] calculated with the Vector Fields program TOSCA [14].

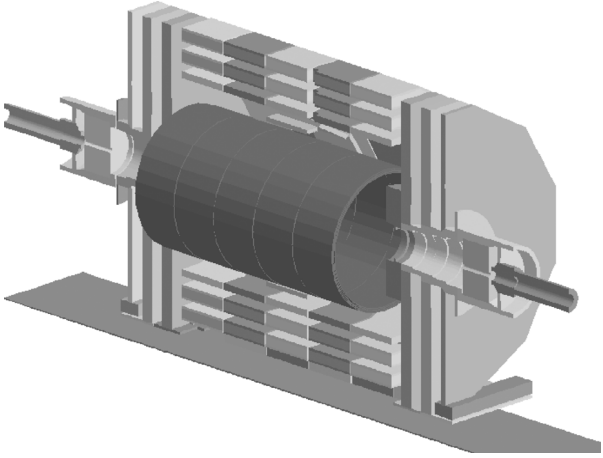

Fig. 1. 3-D model of the CMS magnetic system with a half of the return yoke.

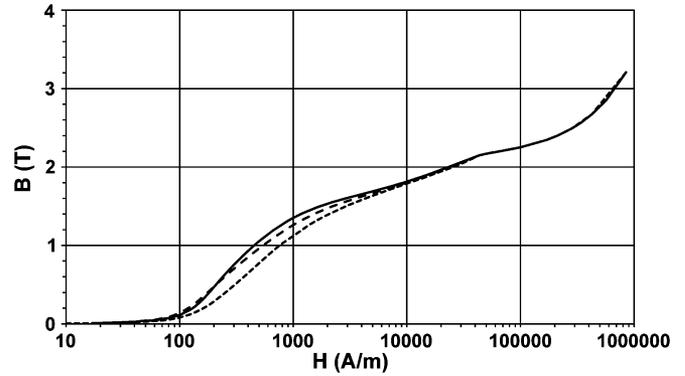

Fig. 2. The BH-curves of the construction steel used in the yoke: smooth line corresponds to thick plates of L2 and L3; long dashed line corresponds to TC and L1 blocks, thin plates of L2 and L3, and connection brackets between the barrel layers; short dashed line corresponds to the nose and end-cap disks and other parts of the return yoke.

In the present paper we describe the CMS field map prepared with the TOSCA CMS model and compare the calculated magnetic field values with the performed measurements.

## II. THE CMS MAGNET MODEL DESCRIPTION

The CMS magnetic system has to be modeled at the entire length and at full range of the azimuth angle: the winding of the CMS coil is not symmetrical with respect to the transverse middle plane, and the return yoke is not symmetrical with respect to either horizontal or vertical longitudinal planes. The coordinate system used in the model corresponds to the CMS reference frame where the X-axis is directed in horizontal plane toward the LHC center, the Y-axis is pointing up, and the Z-axis coincides with the superconducting coil axis. The azimuth angle is counting from X to Y-axis. To describe the full volume of the magnet yoke and to satisfy the limited number of the finite elements to be allowed in TOSCA program, the model is subdivided in two parts. One part describes the entire length of the CMS return yoke at positive X-coordinates as shown in Fig. 1; another part corresponds to full length of the yoke at negative X-coordinates. Both parts comprise the modeled superconducting coil.

The CMS return yoke consists of five barrel wheels having a 7 m inscribed outer radius and a 2.536 m width, two nose disks of 2.63 m radius on each side of the coil, three large end-cap disks having a 6.955 m inscribed radius and one small disk of 2.5 m radius on each side of the magnet. The small disks are followed by ferromagnetic parts of the forward hadron calorimeter and shield of the LHC magnets.

Each barrel wheel except the central one comprises three layers of steel—L1, L2, and L3 connected with brackets. The central barrel wheel has an additional inner layer—tail catcher (TC), made of steel and turned by 5 degrees in the azimuth angle with respect to the dodecagonal shape of the barrel wheels. The thickness of the TC blocks is 0.18 m, the thickness of the L1 blocks is 0.295 m, each block of the L2 and L3 has a thickness of 0.63 m and is comprised of a sandwich of two outer 0.09 m thin plates and an inner 0.45 m thick plate made of different steel. The air gap between the tail catcher and the first barrel layer is 0.562 m, the air gap between the first and second barrel layers is 0.44 m, and the air gap between the second and third barrel layers is 0.395 m. All these air gaps are used to install the muon drift tube chambers to register the muon particles.

The barrel wheels are denoted as follows: YB0 is for central wheel, YB $\pm 1$ are for the wheels next to central wheel, and YB $\pm 2$ are for two extreme barrel wheels. The air gaps between YB0 and YB $\pm 1$ are 0.155 m and the air gaps between YB $\pm 1$ and YB $\pm 2$ are 0.125 m. The barrel wheels feet, the connecting brackets between the barrel layers and two chimneys for cryogenic and electrical leads are included into the model.

The thickness of the two first, YE $\pm 1$ and YE $\pm 2$, end-cap disks on each side of the coil is 0.592 m, the thickness of the third disks, YE $\pm 3$, is 0.232 m, and the thickness of the fourth small disks YE $\pm 4$ is 0.075 m. The air gaps between YB $\pm 2$ and YE $\pm 1$ are 0.655 m, the air gaps between YE $\pm 1$ and YE $\pm 2$ are 0.663 m, the air gaps between YE $\pm 2$ and YE $\pm 3$ are 0.668 m, and the air gap between YE $\pm 3$ and YE $\pm 4$ is 0.664 m. All these air gaps are used to install the muon cathode strip chambers.

Both nose disks, YN $\pm 1$, penetrate partially inside the coil free bore. The distance between YN $-1$ and YN $+1$ is 12.678 m taking into account the YE $\pm 1$ and YN $\pm 1$ deformation under the magnetic forces at the CMS magnet full excitation. The end-cap disk carts upper plates of 0.1 m thickness are included into the model.

### A. Steel Magnetic Properties Description

Three different B-H curves shown in Fig. 2 are used in the model to represent the grades of steel used for the construction of the CMS magnet yoke.

The first curve describes the magnetic properties of the L2 and L3 thick plates. Second curve describes the magnetic properties of the L2 and L3 thin plates, the TC and L1 blocks and the connection brackets between the layers. Finally, the third curve describes the magnetic properties of the nose, end-cap disks, disks cart plates, and ferromagnetic parts of the forward hadron calorimeter and the shield of the LHC magnets. All the BH-curves have been obtained by averaging the measurements done for appropriate specimens of each melt of steel used in the yoke.

### B. Coil Description

The CMS coil consists of five modules of 2.5 m long, and has a length of 12.514 m and inner diameter of 6.3196 m. Four layers of superconductor make the coil thickness of 0.2632 m. In the model the coil dimensions are considered at the temperature of $4°K$ and a scale factor of 0.99585 is applied to all the dimensions. The change of shape of the coil under the magnetic forces is also taken into account. Thus, in the model, the mean radii of the superconductor layers in the central coil module CB0 are 3.18504, 3.25017, 3.3153, and 3.38043 m. In the two adjacent coil modules $CB \pm 1$ the mean radii of the superconducting layers are 2 mm less, and in the two outer coil modules $CB \pm 2$ the mean radii are 5 mm less than in the central coil module. There is one missing turn (out of 2180 designed turns) in the most inner layer of the $CB - 2$ module between $Z = -3.76493$ and $Z = -3.8$ m, which is the reason why the magnetic field produced by the CMS coil is slightly asymmetric with respect to the coil middle transverse plane at $Z = 0$ m.

### C. Model Normalization and Boundary Positions

The currents used in the model exactly correspond to the currents used to map the field inside the inner volume of the coil with the fieldmapper at $B_0$ of 2.02, 3.02, 3.52, 3.81, and 4.01 T.

Each part of the model contains 21 conductors and 1,996,848 nodes of finite element mesh. The number of nodes used in the model is near the present limit of $2 \cdot 10^6$ for the TOSCA program. This value does not permit to mesh at very large radius and restricts the length to be meshed. As the mesh boundaries are too close to the return yoke, this forces too much magnetic flux to return through the yoke, causing an increase of flux in the region instrumented with the muon chambers. A good compromise has been found terminating the finite element mesh at a radius of 30 m and a length of 70 m. These mesh parameters provide a realistic integrated magnetic flux in the transverse plane at $Z = 0$ m [15].

## III. COMPARISON OF MODEL AND MEASUREMENTS

### A. Inner Volume of Coil

Comparisons of $B$ calculated with the CMS TOSCA model at $B_0$ of 4.01 T and measured with the fieldmapper B-sensors at radii of 0.092 and 1.724 m, and with NMR probe at radius of 1.724 m are presented in Figs. 3 and 4.

The measurements performed at radius of 0.092 m and shown in Fig. 3 differ in average from the calculated values by $2.1 \pm 2.0$ mT for the B-sensor located on the negative fieldmapper arm and by $1.4 \pm 1.6$ mT for the B-sensor located on the positive arm. For both B-sensors the measurements and calculations are averaged for each Z-coordinate over the full range of azimuth angle. The typical standard deviation of the measurement averaging is $4 \cdot 10^{-5}$ T.

The measurements performed at radius of 1.724 m and shown in Fig. 4 differ in average from the calculated values by $4.0 \pm 1.0$ mT for the NMR probe, by $2.9 \pm 2.2$ mT for B-sensor located on the negative fieldmapper arm, and by $3.5 \pm 1.4$ mT for B-sensor located on the positive arm of the fieldmapper.

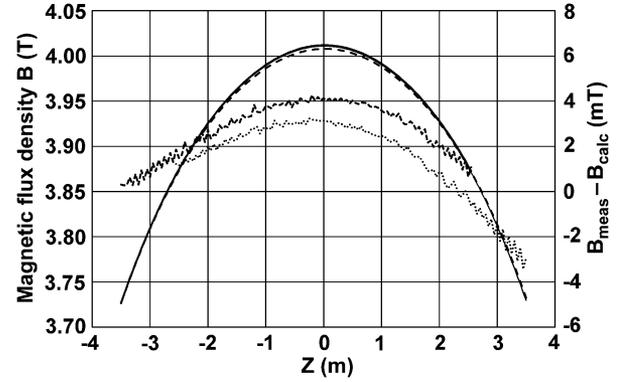

Fig. 3. Comparison of calculated with TOSCA (dashed curve) and measured magnetic flux density (left scale) averaged over the full azimuth angle range. The measurements have been performed with two B-sensors located at radii of 0.092 m with respect to the coil axis on negative (thick smooth curve) and positive (thin smooth curve) fieldmapper arms, respectively. The differences between the measured and calculated values (right scale) are shown by square and round dots, respectively.

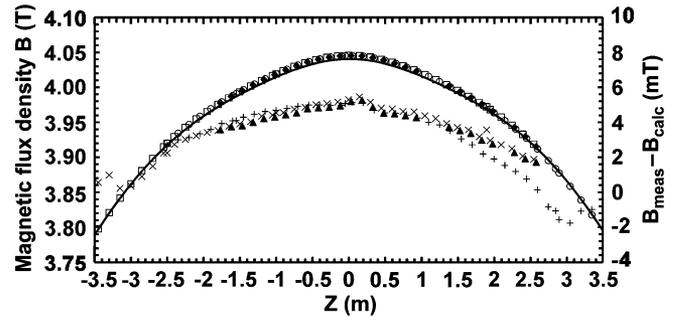

Fig. 4. Magnetic flux density (left scale) measured and calculated along the coil axis in horizontal plane at radius of 1.724 m in the range of $\pm 3.5$ m with respect to the coil transverse middle plane. The measurements are performed with the NMR probe (rhombs), and B-sensor (open squares) located on the fieldmapper negative arm as well as with the B-sensor (open circles) located on the fieldmapper positive arm. Smooth curve represents the calculations done with the CMS TOSCA model. Triangles show the difference (right scale) between the NMR measurements and calculations (the range of measurements is from $-1.767$ to 2.583 m), slanted crosses present the difference between the negative arm B-sensor measurements and calculations, and right crosses display the difference between the positive arm B-sensor measurements and calculations.

Thus, the comparison with the CMS TOSCA model shows excellent agreement with the magnetic flux density measured at $B_0$ of 4.01 T, with discrepancy of 5 mT.

### B. Magnetic Flux Density in Steel Elements

The measurements of the average magnetic flux density inside the steel blocks of the CMS yoke were performed at four $B_0$ values with 22 flux-loops during specially triggered "fast" discharges of the CMS coil. The voltages induced in flux-loops are integrated off-line with time for at least 1300 s. The total uncertainty of the measurements is estimated to be 2% to 7% in different flux-loops. A comparison with the CMS TOSCA model shows that the average ratio of measured over calculated values is 0.97 in the barrel blocks, with a standard deviation of 0.07, and 0.93 in the end-cap disks, with a standard deviation of 0.04. An ad-hoc correction of the CMS TOSCA model predictions in steel blocks of the barrel outer layers L2 and L3 has been implemented using reconstructed tracks from cosmic rays [15].

## IV. MAGNETIC FIELD MAP PREPARATION

To obtain the values of $B$ at each point of the CMS magnetic system in a cylinder of 18 m diameter and 20 m long where steel elements of the return yoke are interleaved with large air gaps, a conception of magnetic and non-magnetic volumes description has been chosen [16], [17]. The volumes represent the groups of the mesh finite elements and are constructed in such a way that their boundaries correspond to field discontinuities, which are due to changes in magnetic permeability of the materials. The field in the volume is continuous and could be found at each point by linear interpolation between the values of $B$-components on a regular grid. To prepare the CMS map at any $B_0$ value, 8640 volumes are used.

To solve efficiently the problem of finding the volume that contains a given point, the volumes are organized in a hierarchical structure and volume finding is reduced to a simple binning problem for each level of the hierarchy. The volumes are grouped in 24 azimuth sectors and searching is made for 360 volumes in the needed azimuth sector only. A simple caching mechanism for the last accessed volume provides about 98% hit rate for both the detector simulation and the event reconstruction reducing the CPU time spent in volume search.

The necessary tables of the three $B$-components are prepared with the OPERA-3d Post-Processor tool [14] using command scripts. The present status of the CMS TOSCA model permits to obtain the field map for any $B_0$ value by simply changing the current in the modeled CMS coil.

The volume approach to the magnetic field access allows the use of algorithms for the field computation in different volumes instead of using a linear interpolation everywhere.

A parameterization of the magnetic field in the region of the CMS inner tracker, where most of the field accesses during the event reconstruction occur, use as parameters the $B_0$ value, the length and the radius of the solenoid. The formulas for axial and radial $B$-components at any radius and Z-coordinate value are expressed as series in terms of the derivatives of azimuthal magnetic flux density at zero radii and the same values of Z-coordinate. For the CMS coil the solenoid radius and the length are considered as formal parameters which are fitted so as to match the parameterized magnetic field as closely as possible with the finite element calculated field map [17].

The parameterization results in 0.1% accuracy on $B$ in the entire inner tracker volume.

## V. CONCLUSION

The CMS magnetic field maps prepared using the TOSCA model meet the accuracy requirements on the magnetic flux density in all regions of the CMS detector except for outer barrel layers where correction of a few per cent is done using reconstructed tracks from cosmic rays.

The magnetic flux density inside the CMS superconducting coil is obtained with a precision of 0.1%.

The Vector Fields program TOSCA is a powerful tool to describe complex heterogeneous magnetic systems with precision as required by present particle detectors.